\documentclass[doublespacing]{elsart_modified}

\usepackage{natbib}

\usepackage{graphicx}

 \usepackage{amssymb}

\usepackage{verbatim}

\begin{document}

\begin{frontmatter}



\title{Formation and Detection of Earth Mass Planets around Low Mass Stars}


\author{Ryan Montgomery} and 
\author{Gregory Laughlin}

\address{University of California Santa Cruz
  Santa Cruz CA 95064 (U.S.A.)}


%
%
%
%
%


\end{frontmatter}



\begin{flushleft}
\vspace{1cm}
Number of pages: \pageref{lastpage} \\
Number of tables: \ref{lasttable}\\
Number of figures: \ref{lastfig}\\
\end{flushleft}


\newcommand{\sep}{; }


\pagebreak

\noindent
\textbf{Proposed Running Head:}\\
  Formation and transit detection of terrestrial planets

\vspace{3cm}
\noindent
\textbf{Please send Editorial Correspondence to:} \\
Ryan Montgomery \\
UC Santa Cruz Astronomy Department\\
1156 High Street - ISB 256\\
Santa Cruz, CA 95064, USA. \\
\\
Email: rmontgom@ucolick.org\\
Phone: (831) 459-3809 

\vfill

\pagebreak


\noindent
\textbf{ABSTRACT}

We investigate an \emph{in-situ} formation scenario
for Earth-mass terrestrial planets in short-period,
potentially habitable orbits around low-mass stars
($M_* < 0.3 M_\odot$). We then investigate the 
feasibility of detecting these Earth-sized planets.
We find that such objects can feasibly be detected 
by a ground-based transit survey if their formation 
frequency is high and if correlated noise can be controlled 
to sub-milli-magnitude levels.
Our simulations of terrestrial planet formation follow
the growth of planetary embryos in an annular region spanning
$0.036 \, \rm{AU} \le a \le 0.4 \, \rm{AU}$ around a fiducial M7 
($0.12 M_\odot$) primary.
Initial distributions of planetary embryos are 
calculated using the semi-analytic evolutionary 
model outlined by \citet{2006Icar..180..496C}.
This model specifies how planetary embryos grow
to the stage where the largest embryo masses lie in the 
$10^{24} g \le M_{embryo} \le 5 \times 10^{26} g$ range
(corresponding to the close of the so-called 
oligarchic growth phase).
We then model the final phases of terrestrial planet
assembly by allowing the embryos to interact with one 
another via a full N-body integration using the 
\texttt{Mercury} code. 
The final planetary system configurations produced in the
simulations generally consist of 3-5 planets with 
masses of order $0.1 - 1.0 M_\oplus$
in or near the habitable zone of the star.
We explore a range of disk masses ($0.2 M_\oplus$ to $3.3 M_\oplus$) 
to illuminate the role disk mass plays in our results.
With a high occurrence fraction or fortunate alignments,
transits by the planet formed in our simulations could
be marginally detected with modest telescopes of
aperture 1m or smaller around the nearest M-dwarf stars.
To obtain a concrete estimate of the detectability of the
planets arising in our simulations, we present a detailed
Monte-Carlo transit detection simulation incorporating 
sky observability, local weather, a target list of around 200
nearby M-dwarfs, and a comprehensive photometric noise model. 
We adopt a baseline 1.5 mmag level of correlated stellar 
noise sampled from the photometry of the planet-bearing 
red dwarf Gl 436. With this noise model we find that
detection of $1 R_\oplus$ planets 
around the local M-dwarfs is challenging for a ground-based
photometric search, but that detection of planets of 
larger radius is a distinct possibility. The detection 
of Earth-sized planets is straightforward, however,
with an all-sky survey by a low-cost satellite mission.
Given a reduced correlated noise level of 0.45 mmag and an 
intermediate planetary ice-mass fraction of planets orbiting 
a target list drawn from the nearest late-type M dwarfs, a 
ground-based photometric search could detect, on average, 
0.3 of these planets within two years and another 0.5 over 
an indefinitely extended search.
A space-based photometric search (similar to the TESS mission)
should discover $\sim 17$ of these Earth-sized planets during
it's two year survey, with an assumed occurrence fraction of 28\%.

\vspace{\fill}

\pagebreak

\section{Introduction}

To date, planets have been detected in every region of
mass and orbital parameter space to which observational
searches are sensitive. 
In recent years, the minimum detected planetary mass
has been decreasing at a very rapid pace. If one projects
this trend forward, one sees that detection of the first near Earth-mass
extrasolar planets orbiting main sequence stars should 
occur sometime around 2011. Thus, while it has 
always been of great interest to search for terrestrial 
mass planets, the time when success might be obtained now seems 
close at hand. Indeed, both radial velocity observations and 
photometric transit surveys have the capability of achieving the 
detection of Earth-mass planets orbiting nearby low-mass stars.
\citep[see e.g.][]{2008ApJ...679.1582G,2008PASP..120..317N}\\
In the standard paradigm of Solar system formation, the 
terrestrial planets emerge during the later stages of proto-planetary
disk evolution through the coagulation of smaller bodies 
\citep{Safronov_book} \citep[see][for a detailed review]{1993ARA&A..31..129L}.
The initial formation process proceeds rapidly 
\citep[e.g.][]{1993Icar..106..190W,1997Icar..128..429W,1998Icar..131..171K,2000Icar..143...15K}
with a few of the small bodies accreting more matter than their 
companions, thereby increasing their gravitational cross-sections and
experiencing even more rapid growth \citep[e.g.][]{1989Icar...77..330W}
This ``runaway growth'' phase rapidly produces
planetary embryos from the sea of planetesimals. 
As the embryos grow, they perturb the orbits of local 
planetesimals, thus reducing gravitational focusing and 
slowing the continued growth of the planetary embryos.
Nevertheless, the largest planetary embryos still 
accrete material more rapidly than any remaining 
low-mass planetesimals - leading to the so-called
``oligarchic'' growth phase. Eventually, an emergent 
distribution of hundreds of planetary embryos interact 
and collide stochastically, until all the planetary 
embryos have either been incorporated into the forming 
planets or ejected from the system.\\
N-body simulations of the final accretion phases 
of terrestrial planet assembly are able to recreate 
the broad-brush aspects of the inner solar system fairly well 
\citep[e.g.][]{2006Icar..184...39O}.
In particular, the orbital properties of Earth and Venus
analogs are often reproduced, whereas Mars-like and Mercury-like
objects are harder to account for.
Ultimately, however, one would like to test the predictive 
outcome of the current paradigmatic model on systems other 
than our own. It has been shown, for example, that 
initial distributions of lunar-sized embryos similar to 
the distributions that can generate 
the solar system terrestrial planets predict 
Earth-mass worlds orbiting both Alpha Cen A and B 
\citep{2002ApJ...576..982Q,2008ApJ...679.1582G}.
The accretion of such embryos, however, may be problematic
\citep{2008MNRAS.388.1528T,2009MNRAS.393L..21T}.\\
Accretion simulations can also be extended to investigate the formation 
of planets around the low-mass stars that constitute the bulk of 
the local stellar population. For example, \citet{2007ApJ...669..606R}
present a suite of accretion calculations which predict that 
Earth-sized terrestrial planets should be rare around M-type stars 
with $M_* \lesssim 0.3 M_\odot$. 
Their models start with initial planetesimal surface densities that 
scale generally downward from the minimum mass solar nebula-based 
models studied by \citet{2001Icar..152..205C}.
Indeed, the \citet{2007ApJ...669..606R}
result can be encapsulated by stating that low-mass 
planetesimal disks produce low-mass planets. Their work makes it
clear that if the protoplanetary disks associated with the lowest 
mass stars are scaled down versions of the Sun's protoplanetary 
disks, then habitable planets orbiting red-dwarfs will be very rare.
The Raymond et al. result emphasizes the importance of observations probing
inner disk masses as the available mass in these regions is pivotal 
to the formation of habitable planets \emph{in-situ}.\\
Planetary formation around M-dwarfs is now the focus of 
considerable observational and theoretical work. 
One conclusion that seems robust is that Jovian-mass 
gas giant planets around M-dwarfs should be rare if 
outer disk masses scale with stellar mass
\citep[e.g.][]{2004ApJ...612L..73L}.
This prediction is a natural consequence 
of the core accretion theory for giant planets 
\citep[see, e.g.][]{1996Icar..124...62P,2005Icar..179..415H},
and now has supported from observations 
\citep{2004ApJ...617..580B,2006ApJ...649..436E,2006Natur.441..305L}.
Furthermore, ice-giant planets akin to Uranus and Neptune 
should be common around M dwarfs. This result is lent plausibility
by the radial velocity observations of \citet{2008arXiv0806.4587M}
who estimate the fraction of FGK stars with ice-giant
planets to be at least $30\%$, and by microlensing results 
\citep[e.g.][]{2009arXiv0902.1761B}
which give direct evidence of such planets.
Since accretion models are so dependent on the available
mass, the seeming plethora of planets is hard to understand
if the only available initial condition is a scaled 
version of the MMSN. Thus, because M stars 
allow for ready detection of companions down into 
the terrestrial mass range, we think it is of interest 
to investigate the observable consequences of 
a broad range of possible formation scenarios.\\
The plan for this paper is as follows.
In \S 2, we describe the details of our terrestrial 
planet formation model for low-mass stars. This model is
designed from the outset to favor the production 
of habitable planets orbiting red dwarfs. In \S 3, we 
describe the specific simulation outcomes for our model,
and aggregate their statistical properties. 
In \S 4, we construct end-to-end Monte-Carlo 
simulations that realistically assess the near-term 
prospects for finding these planets using inexpensive 
photometric surveys that target nearby stars.
In \S 5, we conclude with our finding that a ground-based
survey with small dedicated telescopes has difficulty providing
significant constraints on our scenario, but that an inexpensive
microsatellite survey can provide a significant test.
\section{Formation  Simulation Method}
We have carried out 40 N-body accretion simulations of 
terrestrial planet formation around low mass stars.
The simulations use initial conditions that are generated with 
a semi-analytic model that is very similar to that described in 
\citet{2006Icar..180..496C}. 
The semi-analytic evolutionary model is intended to encapsulate
the growth of bodies up to and including the oligarchic growth
phase, and includes (1) a disk of equal-mass planetesimals with 
their equilibrium eccentricities and inclinations, 
(2) a gas disk which acts to damp these eccentricities and 
inclinations, and (3) seed planetary embryos which accrete material
throughout the simulation and dynamically heat the planetesimal 
distribution. As the planetary embryos grow, they accrete and 
retain atmospheres, which increase their collisional cross-sections, 
and further accelerate their growth. A semi-analytic evolution 
is deemed complete when the distribution of bodies in the model 
reaches 50 planetary embryos (each separated by 10 mutual Hill Radii). 
At this point the mass in planetary embryos and planetesimals is
approximately equal ($\sim 1.6 M_\oplus$ in each population),
and we take the orbital parameters of the embryos and the surface 
density profile of the remaining planetesimals and fragments as 
the initial condition and transfer the system to a 3 dimensional 
N-body simulation. Our N-body simulations begin with the 50 
planetary embryos and include 50 additional particles as
representatives of the remnant planetesimal population. 
Because we begin with equal numbers, the masses of the
`super-planetesimals' and the planetary-embryos are 
approximately equal (although as the simulations proceed, 
the embryos rapidly outpace the planetesimals).\\
The bodies orbit a late type M-dwarf star (M7, 
$M_\star = 0.12 \ M_\odot$) and are endowed with random 
orbital eccentricities in the range [$0.00 < e_i < 0.01$], 
and random orbital inclinations in the range [$-0.5^\circ < i_i 
< 0.5^\circ$]. The initial longitudes of periastron, $\varpi_i$, 
longitudes of the ascending nodes, $\Omega_i $, and mean 
anomalies, $M_i$, of the bodies are chosen randomly.\\
Due to the uncertainties in the surface density profile
and to facilitate comparison with previous work,
our semi-analytic model draws on an initial solid surface 
density profile chosen to resemble the Minimum Mass Solar Nebula 
\citep{1977Ap&SS..51..153W} in that it falls as $\Sigma = \sigma_0 
\cdot \left( a / 1 \, \rm{AU} \right)^{-3/2}$
until it reaches the 150 K ice-sublimation temperature 
\citep{2004M&PS...39.1859P}
where we truncate the outer edge of the disk. 
This choice of surface density profile plays a
significant role in our investigation - given 
a shallower profile, formation of terrestrial 
planets in M-dwarf habitable zones \emph{in-situ}
becomes more difficult \citep{2007ApJ...669..606R},
and under these conditions, the smaller 
planetary population would be undetectable
given current ground-based photometric precision.
In our model, spurred on the results of 
\citet{2008arXiv0806.4587M} and \citet{2009arXiv0902.1761B}
we imagine that an ice-rich (and up to Neptune-mass) object 
has already formed in the region exterior to the 
ice-line due to the larger initial solid surface 
densities there (in analogy with the formation 
of Jupiter prior to the formation of the Earth).
Given our fiducial luminosity $L_* = 0.02 L_\odot$, 
the $150 \, \rm{K}$ fiducial ice-line lies at 
$r_{150K} = 2.7 \, \rm{AU} \sqrt{L_*/L_\odot} = 0.4 \, \rm{AU}$.
We note that this is only an approximate boundary.
In order to account for the destructive nature of collisions 
in the inner regions (where collisional speeds are faster) 
we truncate the inner edge of the disk by taking 
Chambers' (2001) inner disk Keplerian cutoff velocity:
$v = 54 \, \rm{km/s}$ at $r_{in} = 0.3 \, \rm{AU}$.
Around our low-mass primary ($M_* = 0.12 M_\odot$) this velocity
sets the inner edge of the disk to $r_{inner}=0.036 \, \rm{AU}$.
\\

The value of $\sigma_0$ for the minimum mass solar 
nebula is $\sigma_{0,MMSN} = 7 \, \rm{g/cm^2}$.
\citet{2007ApJ...669..606R}
show unambiguously that when the 
MMSN is extrapolated to lower disk masses, the resulting 
terrestrial planets rarely exceed the mass of Mars.
We thus ask: is there any evidence beyond the MMSN to which 
we can look that might suggest higher surface density 
normalizations for the disks orbiting the lowest mass stars?\\
Sub-millimeter flux measurements of low-mass primary systems
\citep[e.g.][]{2007ApJ...671.1800A}
suggest a possible linear correlation between stellar-mass
and disk-mass for low-mass stars,
albeit with a large amount of scatter.
To quantify this effect, we calculate occurrence
fractions of Earth-mass or larger circumstellar disks 
given a mean solar type disk-mass of $0.005 M_\odot \pm 0.5 dex$
and scaling down to our stars' disk masses by assuming 
that the disk-mass scales with the stellar-mass to the 
0th, 1st or 2nd power. For these three scaling possibilities
(0, 1, 2) we find occurrence fractions of (28\%, 0.7\%, 0.001\%),
respectively.\\
Further, there is a scatter of around an order of magnitude
on either side of the MMSN value of $M_{disk} \sim 0.01 M_\odot$
for a given stellar-mass primary. This serves to restrict the 
likely surface-density normalization to between $0.1$ and $10$ 
times the MMSN value of $\sigma_{0,MMSN} = 7 \, \rm{g/cm^2}$.\\
Searching further afield, we can draw clues from other satellite
systems orbiting primaries other than the Sun. First, consider the 
Jovian satellite Io: Jupiter's mass is $0.001 M_\odot$ and Io,
with $P = 1.8$ d, has $M=8.93 \times 10^{25} \, \rm{g}$.
If we spread Io's mass out following the $r^{-3/2}$ surface 
density profile, we find $\Sigma \sim 11,000 \, \rm{g/cm^2}$ 
at Io's orbital radius. Next, consider the GJ 876 system, 
with $M_* = 0.32 M_\odot$ \citep{2005ApJ...634..625R}.
Planet `d' has a $P = 2$ day orbit 
($a = 0.02 \, \rm{AU}$) and $M = 7.5 M_\oplus$. 
If one assumes that planet d formed from material 
extending to $r \sim 0.075 \, \rm{AU}$ and if this 
material is distributed as $\Sigma \propto a^{-3/2}$, 
then we again find $\Sigma \sim 1.1 \times 10^4 \, 
\rm{g/cm^2}$ at the $P = 2$ d orbital radius.\\
Going out on a limb, one could argue that this 
similarity between Io and GJ 876 d implies that 
low-mass primaries can have
$\Sigma \sim 1.1 \times 10^4 \, \rm{g/cm^2}$ at the 
2-day orbital radius. We adopt this surface density 
normalization in our simulations. Certainly, such an 
assumption is, at this point, mere conjecture, but 
by the same token, a large downward extrapolation 
from the MMSN seems equally uncertain.
In our model, if we extrapolate to $1 \, \rm{AU}$, 
one finds $\sigma_0 = 21 \, \rm{g/cm^2}$.
This value, while high, does fall within the 
large range of acceptable surface densities 
inferred from the extant sub-millimeter data.\\
\citet{2006Natur.441..834C} have noted the 
striking similarity in mass ratios between 
the Jovian planets (Jupiter, Saturn, Uranus)
and their regular satellite systems: 
$m_s/M_p \sim 2 \times 10^{-4}$. 
They suggest that this points to a common formation 
scenario within gas-starved circumplanetary disks. 
In our red dwarf context, the presence of an exterior 
ice-giant planet may play a similar role. 
A Neptune-mass planet orbiting a $0.1 M_\odot$ primary 
at these radii could meter the inward flow of gas through 
the gap that it has opened in the protoplanetary disk. 
The details of the inflow are dependent on the details
of the gas flow around the co-rotational resonances
and so is largely undetermined for the present situation.
We hypothesize that the slow metering of gas into the
inner region allows planetesimals to build up without
suffering the catastrophic effects of Type I migration
\citet{2006Natur.441..834C}.
This would lead to a reduced gas-density in the 
interior orbital regions which would significantly 
reduce the effect of Type I migration. 
In the present work we make the assumption that
such a mechanism makes the effects of Type I 
migration negligible over the planet mass range 
studied, but we have not done a detailed analysis
and this scenario would definitely require further
investigation should observations reveal the presence
of a terrestrial-mass planetary population orbiting
nearby M dwarfs.\\
Terrestrial planets that form in a mode analogous
to Jovian satellites would be expected to be relatively 
volatile-poor, while volatile-rich planets (which would
have larger radii at a given mass) would be 
indicative of a planet that migrated inward rather 
than forming \emph{in-situ} 
\citep{2007Sci...318..210G,2008MNRAS.384..663R}.
The lowest-mass M dwarf target stars are nearby, and so 
follow-up radial velocity measurements are possible 
\citep{2005A&A...431.1105B}.
An Earth mass planet in a 30 day orbit around our
fiducial $0.12 M_\odot$ star would elicit a radial
velocity half-amplitude of $K=0.85 \rm{m/s}$ and so
is detectable given current technology 
\citep{2008psa..conf..181L}.
Measurement of the Doppler wobble generates a mass, and 
when combined with the radius from the transit photometry 
could distinguish between the two scenarios.
(This situation could, of course, be complicated by 
the presence of significant gas atmospheres
augmenting the planetary radii 
\citet[e.g.][]{2008ApJ...673.1160A}).
Assuming similar volatile distributions to solar-type
stars, it seems likely that water content levels on
these planets would be similar to those predicted by
\citet{2004Icar..168....1R}. However, for the purposes 
of probing bulk composition, transits can only make
zeroth-order comparisons.
It is interesting to speculate as to the 
water content of these terrestrial planets but,
as yet, the situation is largely unconstrained.\\
Given these considerations, we set the initial 
solid surface density around our M7 star to be 
$\Sigma = 21 \times \left( \frac{a}{1 \, \rm{AU}} 
\right)^{-1.5} \, \rm{g/cm^2}$, where almost all of 
this mass is in the form of a population of 5 km 
($10^{18} \, \rm{g}$) planetesimals. 
The disk scale-height is $H = 0.05 \left( \frac{a}{1 
\, \rm{AU}} \right)^{1.25} \, \rm{AU}$.
We adopt constant-mass planetesimals, in
keeping with \citet{2006Icar..180..496C}.
This assumption greatly simplifies the semi-analytic
framework with the drawback of possibly introducing a
modest overestimation of the embryo growth rates. 
The remainder of the solid surface density ($0.03\%$ by mass) is
allotted to planetary embryos in accordance with the
constraint, derived by \citet{1993Icar..106..210I},
that the onset of oligarchic growth occurs when the 
embryo and planetesimal masses and surface-densities 
satisfy $2 M_{emb} \Sigma_{emb} = m_p \Sigma_p$.
We set the average inter-embryo spacing to be 10 $R_{Hill}$ 
in keeping with \citet{1996Icar..123..180K}.
We also include a gas disk 
which serves to damp the planetesimal's eccentricities 
and inclinations and causes the planetesimals to drift 
inwards via gas drag. We implement the effects of gas
drag using the prescription of \citet{1976PThPh..56.1756A}.
The gas disk begins with an initial surface 
density of 200 times the solid surface density and,
over the course of the simulation, the gas dissipates 
according to $\Sigma_{gas}(t) = \Sigma_{gas}(0) 
\exp(-t/\tau_{neb})$,
where $\tau_{neb} = 2 \, \rm{Myr}$.\\
As the random velocities of the planetesimals rise, collisions
between planetesimals will become destructive, giving rise to many
small collision fragments. Because these fragments are so small,
the gas is very effective at damping out the eccentricities and 
inclinations of these fragments. Thus we treat this population
as having zero eccentricities and inclinations. This 
greatly strengthens gravitational focusing between the embryos 
and the fragments, thereby increasing the growth rate of embryos 
substantially over what would occur without fragmentation.
We also include the effects of the embryo's nascent planetary 
atmospheres. This also tends to increase the embryo's collisional 
cross-sections, further increasing the growth-rates.
With all of these factors included we find that our embyro
growth-rates are a few ($\sim 3$) times faster than the 
equilibrium-regime approximation of \citet{2006Icar..180..496C} 
(eqn. 6), and $\sim 8$ times faster than the estimation given
by \citet{2002ApJ...581..666K} (eqn. 15).\\
We run the semi-analytic model until it produces 50 planetary 
embryos separated by 10 Hill Radii (at which time the embryos 
have grown to sizes between $1 \times 10^{24}$ and $5 \times 10^{26}$ grams). 
We then replace the remaining planetesimal surface distribution
with 50 ``super-planetesimals'' (each with $m_p \simeq 2 
\times 10^{26} \, \rm{g}$). The 50 embryos and the 50 
super-planetesimals are evolved using the \texttt{Mercury} 
integrator package \citep{1997DPS....29.2706C} which employs
a hybrid algorithm that integrates the bodies with a symplectic 
map and switches to Bulirsh-Stoer integration when close 
encounters occur. In this new framework, the embryos are 
self-gravitating while the super-planetesimals are not.
Each N-body simulation is run for 100 Myr to help ensure 
the long term stability of the emergent systems.
The hybrid integrator uses a variable stepsize to meet 
a Bulirsch-Stoer step-wise error tolerance of $10^{-12}$.
The typical energy error at the end of one of our
integrations is $\sim 4 \cdot 10^{-4}$.\\
In order to address the uncertainty in the disk mass 
used we have run simulations with different disk 
surface-density normalizations. If the total disk-mass 
scales as the stellar mass as is implied by \citet{2007ApJ...671.1800A}
and if the disk surface density normalization in our 
region scales as the total disk-mass, then the mean 
normalization for the disk surface-density, 
$\sigma_0 \sim 7\, \rm{g/cm^2} \times 0.12 = 0.84\, \rm{g/cm^2}$.
The resulting inner disk would only contain $0.17 M_\oplus$ 
and thus cannot host detectable terrestrial planets formed 
\emph{in situ}. In order to quantify this situation, we have 
performed simulations of planetary formation and detection 
with normalizations of $\sigma_0 = 0.84\, \rm{g/cm^2}$ and 
$\sigma_0 = 7\, \rm{g/cm^2}$. The resulting detection 
statistics are low to zero, as expected, and are presented for 
completeness in Table 2.
\section{Formation Simulation Results}
Our semi-analytic model seeks to simulate the final, rapid, 
stages of oligarchic growth that follow the runaway growth 
phase. Initial planetary embryo masses are $M_{emb} 
\sim 3 \times 10^{21} \, \rm{g}$. The semi-analytic model
allows the planetary embryos to achieve presumably more
natural mass and spatial distributions, leading to
a more robust set of initial conditions for the 
final, stochastic, phase of growth.\\
In Fig. 1, the thin and thick lines represent beginning 
and end states of the semi-analytic phase of the simulation.
The solid surface density begins almost entirely in the 
form of planetesimals and after 100 years of growth, the 
inner region has become dominated by the planetary embryos. 
At this point, the planetary embryos are large enough that, 
maintaining the average inter-embryo spacing suggested by 
\citet{1998Icar..131..171K},
the surface density profile can support 50 such embryos.
Using simple time scale analysis, we can compare this 
oligarchic growth rate with other calculations.
We find that $\tau_{grow}$ obtained at $a = 0.1 \, \rm{AU}$
around our M7 dwarf is 29 times shorter than that
found in simulations designed to mimic the solar 
system at $a = 1 \, \rm{AU}$ \citep{2006Icar..180..496C}.
Given the approximate nature of timescales, this is 
consistent with the acceleration to be expected
(see \S3.3 for a detailed calculation).\\

Once the semi-analytic model has generated initial conditions, 
we transfer the remaining planetary embryos to the N-body 
simulations. At this transition, the gas is suddenly removed.
Figures 2 and 3 show the evolution of two of the simulations, 
one from Group A (with nothing exterior to the ice-line),
and one from Group B (with a Neptune-mass body on a $0.6 \, \rm{AU}$ 
nearly circular orbit).
The sizes of each of the symbols are proportional to the radius 
of the body.\\
Figure 2 shows the evolution of simulation A-01. 
After 3,000 years, mutual interactions have 
slightly increased the eccentricities of the bodies, and we 
see the rapid buildup into protoplanets of approximately
Mars-mass.
By 0.1 Myr, we see the formation of a few
bodies with $M_{embryo} \simeq 0.7 M_\oplus$,
predominantly in the innermost regions
of the disk where the surface density is high and where
the timescales are short. These massive embryos
rapidly clear out the surrounding regions of lower mass 
planetary embryos through collisions and scattering. 
Over a timescale of order $\tau \sim 1$ Myr,
these larger bodies slowly scatter or accrete
the remaining planetary embryos.
Finally, those larger remaining bodies settle into low 
eccentricity orbits through the ejection or accretion 
of the remaining smaller planetary embryos, producing
a system of planets with masses between Mars' and Earth's.
These remaining planets are in low eccentricity 
orbits separated by, on average, 30 mutual Hill radii.
Thus, while long term stability is not assured for 
these systems, it is quite likely for extremely
long timescales \citep{1996Icar..119..261C}.\\
Figure 3 shows the evolution of simulation B-01. 
(the Group B simulations included a Neptune-mass 
planet at $0.6 \, \rm{AU}$) The external perturber's 
influence is noticeable early on through the 
excitation of eccentricities of several planetary 
embryos around the 2:1 resonance at $0.378 \, \rm{AU}$ 
as well as the increased eccentricities of the entire 
population relative to that seen in the early
evolution of simulation A-01 (Fig. 2). The orbits rapidly 
become excited due to mutual interactions amongst themselves 
and the perturber, resulting in a dynamically warm system.
Within approximately 10,000 years, the system has built up a handful 
of Mars-mass planets through collisions between the planetary embryos.
The growth rate then slows considerably and over the later stages
of the disk's evolution the larger surviving planets eject
most of the other remaining low mass bodies with the help
of the massive perturber.
The surviving planets are on low-eccentricity orbits 
which are separated by, on average, 17 mutual Hill radii.
The overall situation is similar to that of Group A: 
the planetary embryos have been efficiently 
ejected or accreted onto the remaining planetary bodies 
which are between Mars' and Earth's mass and whose orbits are 
located from the star's habitable zone outwards.

An accretion timescale of 10,000 years seems surprisingly 
short, but it is nevertheless a natural consequence of the 
fact that the growth rate scales as
$\tau_{growth} \propto v \cdot \rho^2$.
Although our region of interest ($a \simeq 0.1 \, \rm{AU}$) 
is close to the primary, the star is also less massive 
($M_* = 0.12 M_\odot$) resulting in very similar 
orbital velocities to the solar system ($v \simeq 1.1 v_\oplus$).
For this planet-forming region we can compare the local
density to the solar system; $\rho / \rho_{MMSN} = 4.67$,
giving our growth timescale relative to the solar system as:
$\tau_{growth}/\tau_\oplus \simeq 24$. 
Thus our simulations after 100,000 years of evolution should 
look similar to simulations done of the solar system at 
approximately 2.4 million years \citep{2001Icar..152..205C}.
This is indeed the case.

Considering planets that exceed the mass of Mars by the 
end of our simulations, we see that their mass accretion 
is 50\% complete at $\tau_{50} = 8 \times 10^4$ yr 
and 90\% complete at $\tau_{90} = 10^6$ yr. 
The rapid initial evolution arises from the initial 
flurry of accretion activity in the dense environments.
The presence of an external perturber accelerates this 
timescale significantly: the 50\% and 90\% stages are 
reached at median times of $\tau_{50} = 2 \times 10^4$ 
yr and $\tau_{90} = 0.2 \times 10^6$ yr, respectively.\\

With 20 complete simulations in both groups A and B, 
we can consider the statistics of our systems.
Figure 4 shows the resulting histogram of the planetary masses.
The two distributions are roughly similar over
most of the mass range, implying a concordance in the 
average number and masses of planets. At the lowest 
masses, however, we see the primary effect of an external 
perturber on the surviving populations. 
The ice-giant is very efficient at ejecting low 
mass planetary embryos, thus the final planetary 
mass distribution of Group B is made up of a 
population which is distributed around $0.7 M_\oplus$.
In addition, Group A has a low-mass remnant population that 
survived with few if any collisions and only mild scattering 
events. We wish to isolate the general properties of the two
distinct populations of surviving bodies. We thus draw a
division between the two at $0.05 M_\oplus$. The resulting
averaged quantities of interest are presented in Table 1. 
The effect of inclusion of an ice-giant is that it serves 
to raise eccentricities of the low-mass planetary embryos 
from the simulations leading to ejection or accretion onto 
the star. This decreases the average planetary mass slightly, 
and also limits the growth of the remaining planets to 
the inner disk regions.
Previous work that compared terrestrial planet 
formation with and without giant planets
\citep{2003AJ....125.2692L} found that when giant planets
caused significant mass-loss from the system the resulting
terrestrial planets tended to be smaller and closer in to
their host stars. This is consistent with our findings:
when the ice-giant was included we have significant 
mass-loss and find that our resulting planets are smaller
and closer in to the star.
Figure 5 shows the mass found in each bin of semi-major axis
for Groups A and B. Note that the majority of the massive 
planets are located inside or very close to the star's so-called
habitable zone ($0.03 - 0.08 \, \rm{AU}$) \citep{2007A&A...476.1373S}.
In any case, planetary habitability depends on a variety
of factors, and in general can only be evaluated once
the precise orbit, stellar environment and bulk structure
of a given candidate have been determined.

\section{Detection Simulation}
Our simulations create potentially detectable planets.
Our formation scenario is therefore testable, since a 
M7 ($\sim 0.145 R_\odot$) primary presents a generous 0.4\%
transit depth for a terrestrial ($\sim 1 R_\oplus$) planet.
Furthermore, habitable planets orbiting such stars enjoy a 
relatively high a-priori transit probabilities in comparison
to an Earth-analog orbiting a solar type star.
The a-priori geometric transit probability for a
habitable planet emerging from our simulations is
of order
\begin{equation}
  P_{transit} \simeq 0.0045 \ \frac{1 \, \rm{AU}}{a}
  \left( \frac{R_\star + R_p}{R_\odot} \right)
  \frac{1 + e \cos (\pi/2 - \varpi )}{1 - e^2} \simeq 1\%,
\end{equation}
where $\varpi$ is the longitude of periastron referenced to the 
plane of the sky, and the other symbols have their usual meanings.\\ 
Given a transit depth and a specified detection strategy, 
we can compute the detectability of our model planets.
Figure 6 shows the average detection probability per stellar system 
observed as a function of detection sensitivity threshold.
For example, if the noise can be controlled to allow 
the detection of planets generating central transit depths
deeper than 0.4\%, then each planet-bearing star yields a 
$P \sim 3.2\%$ chance of a detection.
The situation thus looks promising. Unfortunately, however, 
the photometric variability of late-type M-dwarfs complicates 
detection and requires us to carry out a more realistic 
appraisal of the detectability of the planets that form in
our simulations.

Fully convective stars near the bottom of the main sequence
often display significant magnetospheric activity and
accompanying photometric variability \citep{2006A&A...448.1111R}.
This is demonstrated by the variability of M-dwarfs,
e.g. Proxima Centauri, which exhibits X-ray flare activity
with possible attendant photometric variation.
In general, this variability can be modeled as longer-period 
correlated or ``red'' noise. Red noise levels in these stars
slows the noise attenuation of phase-folded photometry from
$1/\sqrt{N_{data}}$ to $\sim 1/\sqrt{N_{transits}}$ 
\citep{2006MNRAS.373..231P}.
The half-life of the red noise in this approximation is 
four times the transit periodicity (days to years) rather than 
four times the photometric cadence (seconds to minutes).
Due to this extremely slow error convergence, the utility
of rapidly taking multiple exposures is greatly reduced.
It is thus likely that a successful observational campaign will 
need to attain cadence and signal-to-noise levels that allow
the detection of a transit to high confidence on the 
basis of a single event. The candidate transit then triggers
a relatively intense follow-up campaign that ideally begins
while the initial event is in progress. The \emph{MEarth} survey
described in \citet{2008PASP..120..317N} provides an
example of this type of detection strategy and the 
calculations that follow are based on their scenario.\\
Our approach requires an estimate of stellar photometric 
noise over a time scale of hours. As a concrete example, 
we take our noise model from photometric 
observations of GJ 436 \citep{2007A&A...472L..13G}
as obtained using the Wise 1m telescope on April 24th, 2007. 
This data set illustrates, to a first approximation, the
sort of data that might be regularly obtained by a ground-based
survey using dedicated small telescopes.
The Gillon et al. photometry was obtained during an 2.9-hour
observation centered on the transit of Gliese 436 b.
Adopting only the photometry outside of transit yields 177 
photometric points over 2.12 hours with a standard deviation 
of 0.39\%. No previous treatment has been applied to the data
aside from the standard reduction and differential photometry.
We process this data to produce a photometric red noise signal 
that can be applied to generate Monte-Carlo data as follows.\\
First, we distribute the \citet{2007A&A...472L..13G}
photometric data into 10-minute bins, 
and use the photometric scatter within these 
bins to calibrate a model for the white noise contribution.
We generate synthetic 177-point datasets of pure Gaussian 
noise and calculate the standard-deviation within bins 
of 14 data points. We repeat this procedure until 
the restricted standard-deviation converges and the 
difference from the known noise amplitude supplies our 
correction factor for the white-noise amplitude giving
$\sigma_{white} = 0.36\%$.
We then use the Scargle periodogram of the remaining 
longer-period noise (the smoothed photometry) at periods 
from 10 minutes to 12 hours to generate synthetic 177-point 
datasets. We sum one hundred cosines of random phases 
with the amplitudes given by the periodogram to the 
white-noise. We then iteratively adjust the amplitude 
of the long-period noise until the overall noise 
amplitude of the synthetic datasets converges to 
that of the 
\citet{2007A&A...472L..13G} photometry.\\
Our estimate of the longer-period red noise,
$\sigma_{red} = 0.15\% $, is intended to be independent of the 
noise introduced by the telescope and is assumed to arise 
from intrinsic stellar variability on short timescales.
Red noise will limit the sizes, and hence the masses of the 
planets to which our observational campaign is sensitive. 
A search strategy that forgoes folding must be able to
detect a transit with sufficient signal-to-noise to 
minimize false-alarms and expensive large-telescope 
follow-up to an acceptable level.
Thus, even with zero white-noise, we are limited to stars that
are small enough to show a transit from one of our planets 
that rises significantly above the red noise level. 
For example, if we demand $4 \sigma$ confidence in the 
reality of a given transit signal, we could survey 
only stars small enough to yield transit depths of 
at least $\sim 0.6\%$. The red noise amplitude will vary with
time, and from star to star. Hence the estimate of 
$\sigma_{red} = 0.15\%$ adopted here should be understood
to provide only a starting point for a more detailed
investigation.\\

Obviously, our strategy benefits from a ready supply of small nearby 
stars. To quantify the available census, we adopt the L{\'e}pine-Shara 
Proper Motion Catalog - North \citep{2005AJ....129.1483L}.
We begin with the local subsample of
the proper motion catalog \citep{2005AJ....130.1680L}
consisting of 4131
dwarfs, subgiants and giants located within 33pc of the Sun.
Following the lead of \citet{2008PASP..120..317N},
we use color cuts to restrict the list to 2401 local 
M-dwarfs by requiring that 
$H - K < 0.7, J - H > 0.12, J - K > 0.7,$ and $8 < H < 15$.
We estimate the stellar mass using the \citet{2000A&A...364..217D}
mass-luminosity relation and then we use the \citet{2006ApJ...651.1155B}
empirical mass-radius relation to approximate the stellar radii. 
Given the stellar radii, we
further restrict the sample to harbor only stars with
radii small enough to allow detectable transit
depths (above a threshold of $TD_{crit} = 4 \sigma = 0.6\% $). 
We thus adopt a transiting planetary radius of $1.2 R_\oplus$,
and require this size planet's transit signal to be equal to 
or greater than $TD_{crit}$.
This adopted planetary radius is chosen so that we will only
exclude from our target lists those stars which are so large
that detection of any of our formed planets is unlikely.
In effect, this restricts our target list to only those 
late-type M-dwarfs of radius $R_* \simeq 0.12 R_\odot$. 
This profoundly limits the sample, leaving a prime catalog
of 169 local, very late-type M-dwarfs. These stars 
offer the single best opportunity to detect a habitable
transiting planet from the ground.

We need to monitor M dwarfs with repeated visits at a 
cadence that is slightly more frequent than the expected
transit duration. We must therefore evaluate the signal-to-noise 
levels for any star-telescope combination as a function of
integration time. Following \citet{2008PASP..120..317N}
we use the bolometric corrections of \citet{2000ApJ...535..965L}
to derive the stellar luminosities.
Luminosities and radii yield values of the 
stellar effective temperatures, $T_{eff}$.
Our mass and radii estimates then provide surface 
gravity ($\log g$) estimates  for our stars.
With $T_{eff}$ and $\log g$, we can specify a 
synthetic stellar spectra, $f(\lambda)$ 
\citep{1999ApJ...512..377H}.
When combined with a target's distance, $d$, 
this synthetic spectra provides an estimate 
of the incident stellar flux
\begin{equation}
  F(\lambda) = f(\lambda) \left( \frac{R_*}{d} \right)^2
  \frac{\lambda}{hc} \, ,
\end{equation}
where $F(\lambda)$ has units of photons/cm$^2$/sec/\AA. 
With these assumptions, the number of photons received,
$N_{ph}$, from a given target in exposure of 
length $t$ on a telescope of diameter $D$, is
\begin{equation}
  N_S{ph} = t \times \pi (D/2)^2 \times 
  \int F(\lambda) T(\lambda) \rm{d} \lambda \ ,
\end{equation}
where $T(\lambda)$ is the telescope's transmission function.
Our transmission function mimics \citet{2008PASP..120..317N}
`i+z' filter that opens at around 700nm and stays open 
for longer wavelengths, combined with the quantum efficiency 
of a typical commercially available CCD 
(which falls redward of 800nm and completely closes off by 1000nm). 
We enforce an additional throughput loss of 50\% to account 
for other unaddressed potential effects.\\
With a photon count, we can obtain an accurate estimate of 
the photometric signal produced by a given target.
We approximate the signal-to-noise ratio of an exposure as:
\begin{equation}
  SNR = \frac{N_{ph}}{
    \sqrt{N_{ph} + 
      N_{ph}^2 (\sigma_{scint}^2 + \sigma_{red}^2) +
      n_{pix} (N_{sky} t_{obs} + N_{dark} t_{obs} + N_{read}^2) }}
\end{equation}
where $\sigma_{red} = 0.15\%$ as stated above, and $\sigma_{scint}$ 
is given by the scintillation expression of \citet{1998PASP..110..610D}.
The quantity $n_{pix}$ represents the expected number of pixels covered 
by the stellar image. $N_{sky} = 10^{-5} \, \rm{e^- pixel^{-1} sec^{-1}}$
is an estimate for the sky brightness. $N_{dark} = 1.4 \cdot 10^{-4}
\, \rm{e^- pixel^{-1} sec^{-1}}$ is an estimate of the 
dark-current noise based on the Nickel 1-meter telescope's
direct imaging camera CCD at Mount Hamilton. 
$N_{read} = 11.7 \, \rm{e^- pixel^{-1}}$ is the amplifier-generated 
readout noise, again based on the Nickel 1-meter's CCD specifications.
For reasonable ground-based exposure times the readout noise 
dominates over the dark-current and sky, contributing $\sim 0.4$ 
mmag noise. Thus, for red noise levels below $0.4$ mmag the 
readout noise of our camera will begin to dominate the SNR value.
For fixed telescope size and exposure time, the required SNR yields a 
tradeoff between spurious transit-like noise and missed transits.
Our goal is to optimize the choices of telescope diameter, 
exposure time, and SNR value for recovery of our simulated
planetary population.
Our exposures must be a few times shorter than the 
expected transit duration. For close-in terrestrial 
planets whose transit durations are less than 60 minutes, 
this limits the exposure time to less than about 20 minutes.
Given this maximum exposure time, we can estimate
$N \sim 21$ observations per 7-hour night. 
Three targets on a given night yields $\sim 7$ 
photometric observations per star.
We thus also require that noise be discernible 
from a transit signal at the $4.3 \sigma$ level,
allowing $\sim 99.8\%$ confidence in any given 
($\sim 7$ point) lightcurve.
We adopt telescope sizes of 1 meter in keeping a realistic 
expectable financial investment for a dedicated array.\\
Figure 7 shows the number of stars that are bright enough
to allow observation within 20-30 minute exposures for 
varying stringencies of the required SNR. 
Note that there are few stars which allow a statistically
significant detection. For example, if we adopt a detection
criterion of $3 \sigma$, the number of available target stars
is unacceptably low ($\sim 10$).
Thus, for correlated noise levels $\sigma_{red} \sim 0.15\%$,
the prospect of detecting transiting Earth-sized planets with 
an array of meter-class ground based telescopes is challenging.
Our Monte-Carlo detection simulations therefore examine 
the effects of different levels of correlated noise.

Our observability simulations extend the work of 
\citet{2003PASP..115.1355S} and specific details of the 
computational procedure can be found in that work.
A Monte Carlo simulation is initialized with the 
planetary populations (drawn from our simulations),
a list of observers, and a stellar target list
(the low-mass stars defined by $TD > TD_{crit}$).
Each observer has an associated location 
(latitude, longitude and altitude) and weather 
(average fraction of clear/cloudy nights per year). 
Each target has an associated position (R.A. and Dec.), 
stellar mass, and a photon flux over the wavelength 
region of interest. 
The Monte Carlo routine starts by assigning which 
targets will host transiting planets in the simulation. 
The program assigns planetary systems (the end-state 
of our N-body simulations) to a fraction of these 
target stars and then orients those planetary systems 
randomly with respect to the line-of-sight to Earth.
Assigned planets maintain their orbital periods, 
and hence assume a semi-major axis in concordance with
the parent star mass and Kepler's third law.
The simulation proceeds through an observing campaign
night by night. Every night, the simulation must 
determine where the weather is favorable: while
season-based weather patterns have not been figured into the
model, the fraction of clear and cloudy nights at each observing
location is known\footnote{http://www.ncdc.noaa.gov/oa/climate/online/ccd/cldy.html}.
The night's weather for each location is determined by drawing
from these probabilities.
If clear or cloudy, all observers common to the location will
be affected identically. If the weather is determined to be partly
cloudy in a given location, then it is possible that some observers
at that location will be able to observe while others will not.
For each observer at a location with favorable weather,
the Julian dates (JD) of sunrise and sunset
at the observer's location are calculated for the current date in
the campaign. The JD of sunset, the position of each target,
and the observer's location are used to calculate air masses for
each target in the target list. Any targets that pass the airmass
cutoff ($\sec (z) = 2.5$ in our simulations) at sunset are then checked to
ensure they will pass the airmass limit for at least 2 hr.
Each observer will then construct a target-list from those targets
that have passed the airmass cutoffs and whose required exposure
times are short enough to allow multiple targets to be observed every hour.
The observer then cycles through this observing list every hour until
one of the targets fails an airmass test, at which point it is dropped
and a new target is added to the observing list. 
On each visit, an observer integrates long enough to 
reach a $5 \sigma$ signal-to-noise ratio before slewing 
to the next target (1 minute of combined slew and readout 
time is added between each target in the observing schedule). 
This level of signal-to-noise is higher than that used to 
signal a detection so that we are able to detect transits
more quickly and reliably. Once a detection has been made,
we anticipate extensive follow up observations by larger
telescopes, however in this preliminary study we do not modify 
our observing plan as a result of transit detections.
The list is repeated every hour so as to catch the majority 
of transits that might occur over the course of the night.
In the event that one of the targets hosts planets, and that one of
the planets happens to transit during the night's observations,
the line-of-sight separation between the planet and its host star
is calculated. Using this separation, as well as
the radii of the transiting planet and its host star, the
photometric transit depth is determined using the approximate
formulae of \citet{2005ApJ...622.1118O}.
This process is repeated for each observer in the 
observer list, and the aggregate process constitutes 
1 night of the campaign. The simulation then increments 
its internal calendar by 1 day and repeats the 
procedure for the overall duration of the campaign.
Using this rotating observation method rather than 
continual monitoring of individual targets allows us
to monitor several stars simultaneously with only
minimal loss of coverage: the typical transit length
is one hour, so one observation every
hour should catch any transits that occur. 
In comparison, the typical orbital period is between 10
and 60 days, so if we wished to monitor individual
stars continually we should do it at least over this
time range. This severely limits the rate at which we 
could study our targets, extending the necessary time 
to discovery by many years.

The results of the Monte-Carlo simulation depend
on the number and distribution of
observers, the stringency of the detection criterion (SNR-value),
and the adopted level of correlated noise.
We set the fraction of stars with planets to 100\%, 
this result is of course downwardly scalable.
As an illustrative case of a viable search strategy,
we distribute 10 observers in longitude around the Northern hemisphere. 
We choose this distribution of observer locations envisioning
the utilization of a pre-existing infrastructure of small 
telescope observers \citep{2003PASP..115.1355S}, 
and we note that such a distribution could be altered 
more observers at fewer sites without a significant impact 
on the cadence or quality of the results.
We choose this distribution of observer locations envisioning
the utilization of a pre-existing infrastructure of small 
telescope observers \citep{2003PASP..115.1355S}, 
and we note that such a distribution could be altered 
more observers at fewer sites without a significant impact 
on the cadence or quality of the results.
Each observer operates a 1-meter telescope and signals a potential 
transit whenever a photometric measurement falls $4.3 \sigma$
below the night's average. This stringent detection criterion
of $4.3 \sigma$ was chosen to reduce false-positive occurrence to
acceptable levels. Due to the highly non-Gaussian nature of
the false-positive rate, we iterated to determine this optimal
value of $4.3 \sigma$ where we were able to reduce the
false-positive rate to (on average) 19 over the entire two-year
campaign while still allowing reasonable integration times.
We vary the amplitude of the correlated red noise from 
1.5 mmag to zero, and for each red noise level we repeat 
the model observing procedure several (5-10) times.
The averaged results stemming from 135 Monte-Carlo simulations
are presented in Fig. 8. A ground-based transit search 
of the above specifications is able to detect Earth-sized 
planets with acceptable levels of false-positives when red 
noise is controlled to the level of 0.45 mmag (30\% of the 
adopted value). Over the course of a two year campaign the observers
produce approximately 9000 photometric light-curves. Of these 
light-curves, on average 5.5 contained true transits and 1.3 of these
were detected at the reduced level of correlated noise (0.45 mmag).
At the end of these campaigns there are another 1.5 transiting 
planets that could be detected but haven't yet, thus a campaign 
that utilized more observers or an extended search would be 
expected to detect a total of $\sim 2.8$ transiting planets 
given the reduced level of red noise. 
If we utilize the $h=0$ scaling occurrence fraction of 
$f=28\%$. We obtain an estimate of $0.8$ detections.
Our results are tabulated
based on planetary radii drawn from the results of 
\citet{2007ApJ...665.1413V} 
that assume ice-mass fractions of 0\%, 50\%, and 100\%.
This allows us to cover a reasonably expected range
of possible planet compositions and sizes, the resulting
planetary radii, number of planets detected under said
observing campaign and the number of potentially detectable
planets under this observing configuration are presented in
Table 2. These other simulated observing campaigns are only 
different in the planetary radii used, all other parameters
were held constant to facilitate comparison.
We find that the number of detections is small regardless
of the ice-mass fraction of the planets, similarly the
number of potential detections is uniformly small,
making a statistically significant null-result elusive.
Irrespective of the red noise level, given our SNR detection
threshold of $4.3 \sigma$, there were an average of 19 spurious
false-positive `transits' reported over each of the 2-year 
observing campaigns.\\

Transit detection from space has been a topic of discussion 
following the direct detection of HD 209458b by the 
Spizter Space Telescope \citep{2005Natur.434..740D}, 
as well as the constraint on its albedo by the 
MOST telescope \citep{2006ApJ...646.1241R}.
To examine the potential detection of our simulated 
planets using a dedicated space-based mission, 
we ran a simulation meant to emulate the general 
characteristics of the proposed TESS satellite. 
As before, we use our suite of planet-forming simulations
to populate the target stars. 
The TESS satellite is designed to have six small 
($\sim 13 \rm{cm}$) wide-field cameras observing stars in 
the direction of the anti-solar point, with a total field of 
view of $\sim 2000$ square degrees.
This setup will yield $\sim 36$ days of continuous photometry 
for each star over it's two year survey.
Using a simplified version of our Monte-Carlo code, 
with a fixed exposure time of 1 minutes, we simulated
observations of the 1385 best candidate stars in the
LSPM-North catalog. This sample of stars was determined
as described above but with the stellar size cut 
($TD_{crit}$) based on the reduced noise level due to 
the lack of atmospheric scintillation.
We then analyzed the resulting lightcurves by sliding a
characteristic transit template along the lightcurve and for 
each time-alignment calculating a $\chi^2$ value between the 
lightcurve and the template. We then divide this $\chi^2$ 
value by the $\chi^2$ of a null hypothesis (no transit).
We fold this figure of merit time-sequence along the
possible planetary periods, and search for periodic signals.
Of the 144 stars that had transits, 31 were clearly detectable
while the rest are too small relative to the stellar 
(red) noise to be detected. If we assume that the Southern
sky would allow another 31 detections and use an
occurrence fraction of 28\%, we find that such a mission
would allow the detection of $\sim 17$ of the planets 
in our simulations after the two year survey.

\section{Summary}
It is generally agreed that if terrestrial planet 
formation around low-mass M-dwarfs is a scaled down 
version of solar system formation, then planets in 
the habitable zones of low-mass red dwarfs
will be small, dry, and thoroughly inhospitable
\citep[see, e.g.][]{2007ApJ...669..606R,2007ApJ...660L.149L}.
Nature, on the other hand, tends to defy expectations.
Our tack here has been to investigate an alternative,
optimistic scenario for terrestrial planet formation
around nearby low mass stars, which assumes that 
disk-mass is largely independent of stellar mass.
In this scenario, the planets that emerge are more
akin to the Jovian satellites rather than Mercury, 
Venus, Earth and Mars. Our motivation for this 
investigation stems from the fact that our hypothesis 
is potentially testable, and at a relatively low cost.\\
Our approach couples end-to-end simulations of both 
the \emph{in-situ} formation and the transit-detection 
of terrestrial mass planets around local, 
late-type M-dwarf stars.
Our formation simulations followed the 
accretional growth of planetary embryos,
around an $M_* = 0.12 \, M_\odot$ star, from 
the onset of oligarchic growth ($M_{emb} \sim 
3 \times 10^{21} \, \rm{g}$) 
using a semi-analytic method, through the stochastic 
late-phases of growth using a full 3-D Nbody simulation. 
These simulations resulted in systems of several (3-5) 
planets with masses comparable to Earth in stable orbits 
in or near the star's habitable zone as well as 1-2 
planets per system with masses comparable to Mars.
We found that these results were largely insensitive 
to the presence or absence of an exterior perturbing 
Neptune-mass planet, but detectable planets are more 
likely to occur if an external perturber is present.\\
Photometric red noise makes transit detection of these 
small planets challenging.
We have explored the impact of varying levels of 
red noise on the potential utility of dedicated 
ground-based telescope arrays.
Our method consisted of a detailed Monte-Carlo 
transit search code implementing realistic weather, 
stellar flux, telescope and M-dwarf noise models.
We find that a formation frequency of $\sim 28\%$ and 
nightly stellar noise levels in the vicinity of $\sigma 
\sim 0.45$ mmag allow a longitudinally distributed 
array of ten dedicated meter-class telescopes carrying 
out a targeted search to detect, on average, 0.3 transiting 
terrestrial planets within two years of operation and
another 0.5 potentially detectable planets.
However, a satellite based survey similar to TESS
is capable of detecting $\sim 17$ planets after 
it's two year survey.\\
Time and again over the past decade, nature has 
surprised planet hunters with the sheer diversity 
of planetary systems. Experience has shown that 
planet formation under a variety of conditions can 
be more efficient than the example of our own 
system would suggest. Our approach in this paper 
has been to investigate a planet building model 
that lies at the optimistic, yet still justifiable 
range of formation scenarios. Should our picture 
prove correct, then the prospects for discovery of 
truly earthlike, alarmingly nearby, and potentially 
habitable planets may lie close at hand.

\ack
This research has been supported by NSF CAREER Award AST-0449986, 
and by the NASA Origins of Solar Systems Program through grant
NNG04GN30G.\\
The authors wish to thank David Charbonneau and John Chambers 
for helpful discussions and technical advice concerning the 
\texttt{Mercury} code.
Also thanks to Scott Seagroves and Justin Harker for the use 
of and advice concerning the \texttt{Transit-Search} code.\\
The IDL astronomy user's library is available at 
http://idlastro.gsfc.nasa.gov\\

\label{lastpage}

\bibliography{mybib.bib}
\bibliographystyle{elsart-harv}


\clearpage	

\begin{table}
\begin{center}
\textbf{Final Planetary Configuration Statistics}
\begin{tabular}{ll|c|c||c|c}
\hline
\hline
\multicolumn{2}{c|}{Simulation Group} & 
$\overline{N}_{Planet}$ & $\overline{M}_{Planet}$ & Mass Lost &
$\overline{N}_p$ \\
\hline
 A & No Ice-giant       & 4.4 & 0.79 & 2\%  & 1.4 \\
 B & Ice-giant present  & 3.8 & 0.68 & 18\% & 0.2 \\
 \hline
\end{tabular}
\caption[NperSys] {
	\label{MercAverages}
        ``Planet'' indicates masses above $0.05 M_\oplus$.
        $\overline{N}_{Planet}$ is the average number of planets per system. 
        $\overline{M}_{Planet}$ is the median mass of planets per system.
        $\overline{N}_p$ is the average number
        of planetary embryos that survived per system.
	}
\end{center}
\end{table}

\begin{table}
\begin{center}
\textbf{Comparison of Planetary Compositions, Radii and Detectability}
\begin{tabular}{c|c|c|c}
\hline
\hline
$Ice$ & $\overline{R}_p$ & $\overline{N}_{detect}$ & $\overline{N}_{possible}$\\
\hline
100\% & 1.46 & 0.4 & 0.6\\
50\%  & 1.20 & 0.3 & 0.5\\
0\%   & 0.93 & 0.1 & 0.3\\
\hline
50\%*  & 0.68 & 0.1 & 0.1\\
50\%** & 0.24 & 0.0 & 0.0\\
 \hline
\end{tabular}
\caption[NperSys] {
	\label{Composition}
        \label{lasttable}
        Columns indicate the planetary ice-mass-fraction
        (Valencia et al., 2007), median resulting planetary radius (in
        Earth radii), the number of planets detected in our observing 
        scenario after the two year simulated observing campaign, 
        and the number of additional planets potentially detectable 
        given the levels of rednoise present, respectively. 
        Assuming an occurrence fraction of 28\%.\\
        The bottom two rows respresent our low disk-mass simulations
        with normalizations of * $7\, \rm{g/cm^2}$ and 
        ** $0.84\, \rm{g/cm^2}$.
	}
\end{center}
\end{table}

\clearpage


\begin{figure}[p!]
  \begin{center}
    \includegraphics[width=4in]{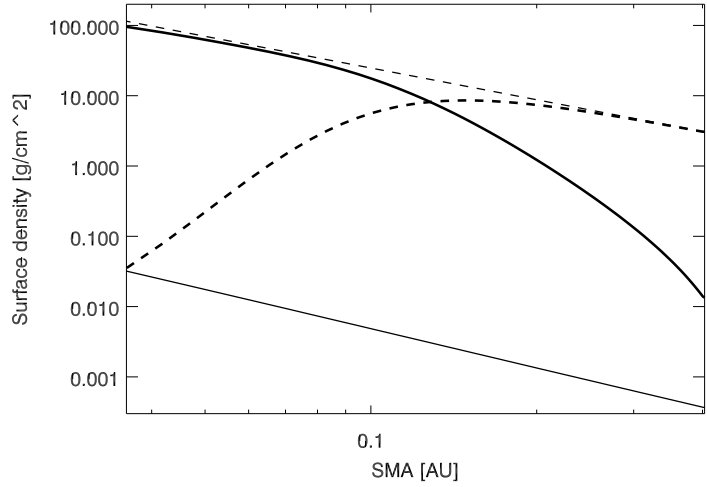}
    \caption[SAM growth]{
      \label{SAM_growth}
      Beginning/end (thin/thick lines) surface density profiles
      of the planetary embryos and planetesimals (solid and dashed 
      lines, respectively) from the semi-analytic model.
      The semi-analytic model runs until $\sim 150$ years.
    }
  \end{center}
\end{figure}

\begin{figure}[p!]
  \begin{center}
    \includegraphics[width=4in]{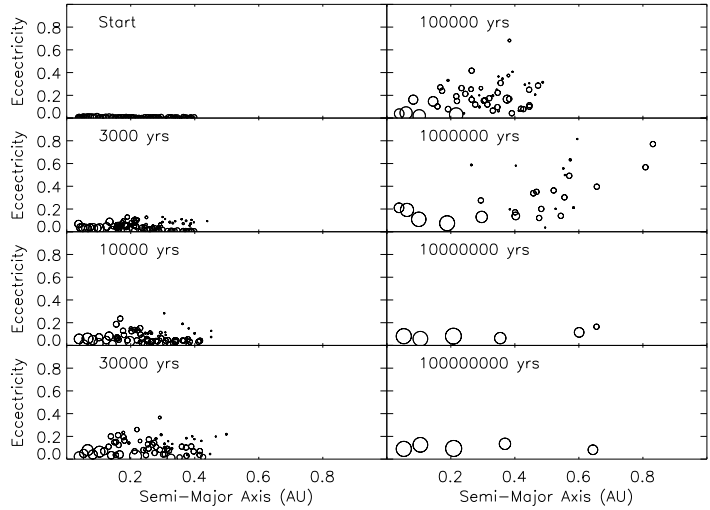}
    \caption[Sample Evolution A-01]{
      \label{No_Evo}
      Evolution of Simulation A-01 (no ice-giant present)
      ending with planets of mass 0.86, 0.79, 1.03, 0.36 and 0.22 $M_\oplus$
      (in order of increasing semi-major axis) after $10^8$ years. 
      The size of the symbol is proportional to the radius of the body.
    }
  \end{center}
\end{figure}

\begin{figure}[p!]
  \begin{center}
    \includegraphics[width=4in]{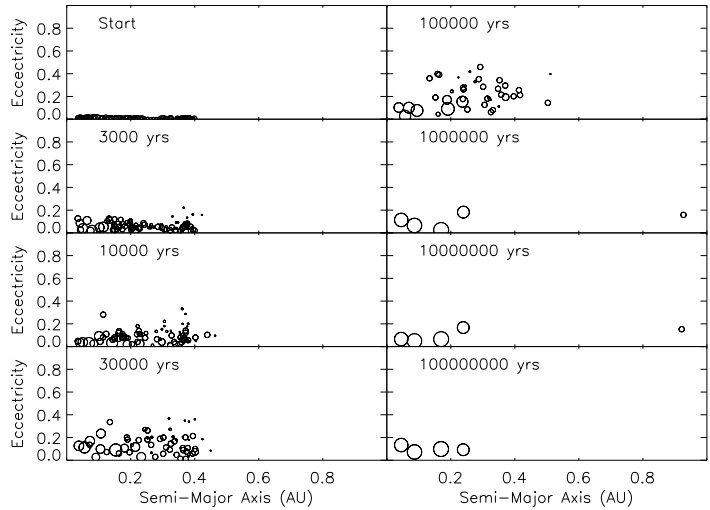}
    \caption[Sample Evolution B-01]{
      \label{IG_Evo}
      Evolution of Simulation B-01 (ice-giant at 0.6AU)
      ending with planets of mass 0.60, 0.75, 0.83, and 0.41 $M_\oplus$
      (in order of increasing semi-major axis) after $10^8$ years.
      The size of the symbol is proportional to the radius of the body.
    }
  \end{center}
\end{figure}

\begin{figure}[p!]
  \begin{center}
    \includegraphics[width=4in]{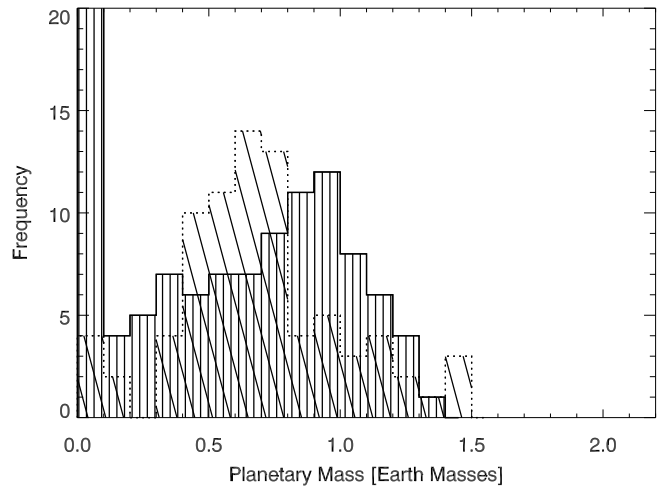}
    \caption[Mass Histogram]{
      \label{mplanets}
      Final planet masses using a non-uniform initial mass distribution.
      Group A is represented by the solid line (no ice-giant present).
      Group B is represented by the dotted line (ice-giant at 0.6AU).
    }
  \end{center}
\end{figure}

\begin{figure}[p!]
  \begin{center}
    \includegraphics[width=4in]{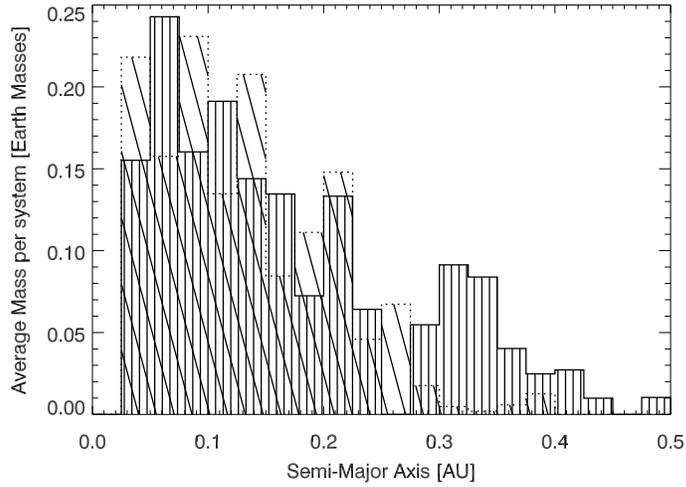}
    \caption[Spatial Distribution of Mass]{
      \label{weighted}
      The amount of mass located in each semi-major axis bin.
      Group A is represented by the solid line (no ice-giant present).
      Group B is represented by the dotted line (ice-giant at 0.6AU).
    }
  \end{center}
\end{figure}

\begin{figure}[p!]
  \begin{center}
    \includegraphics[width=4in]{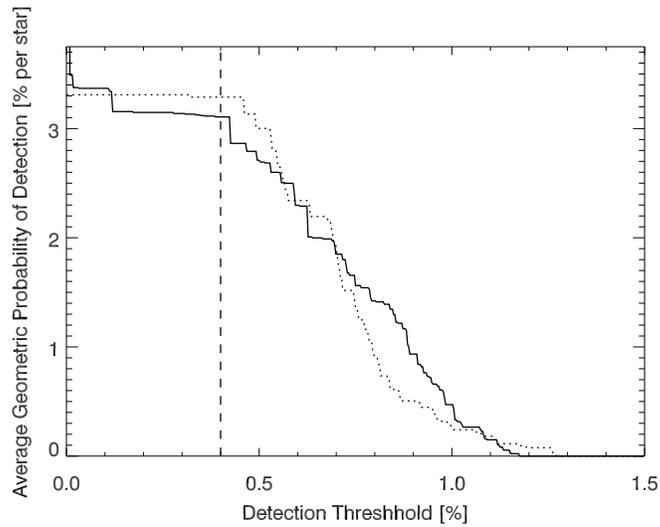}
    \caption[Geometric Transit Probability]{
      \label{transprob}
      Probability of detecting a transit of a planet
      around its host star as a function of the telescope's
      sensitivity. Dotted line shows the simulations with 
      an ice-giant, solid line shows the simulations without.
      An approximate sensitivity of sub-meter class telescopes 
      is shown for comparison ($\sim 3$ milli-magnitudes).
    }
  \end{center}
\end{figure}

\begin{figure}[p!]
  \begin{center}
    \includegraphics[width=4in]{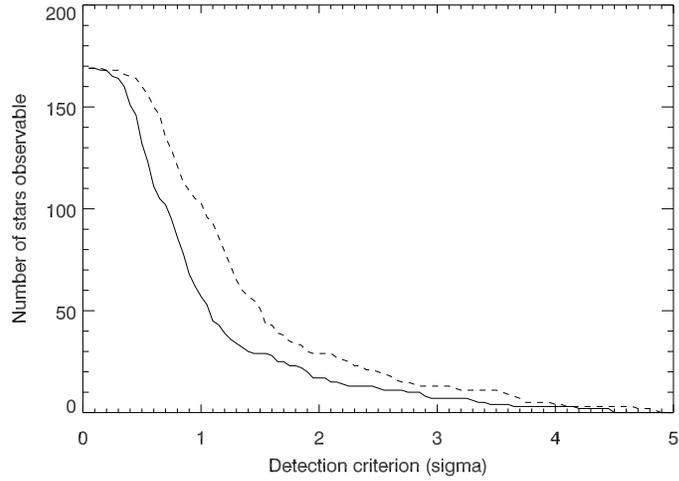}
    \caption[Recovery vs Aperture]{
      \label{RecVsAper}
      Number of stars which could yield detectable transit signals 
      in a 20 (30) minute exposure as a function of signal-to-noise 
      requirement, shown by the solid (dashed) line.
    }
  \end{center}
\end{figure}

\begin{figure}[p!]
  \begin{center}
    \includegraphics[width=4in]{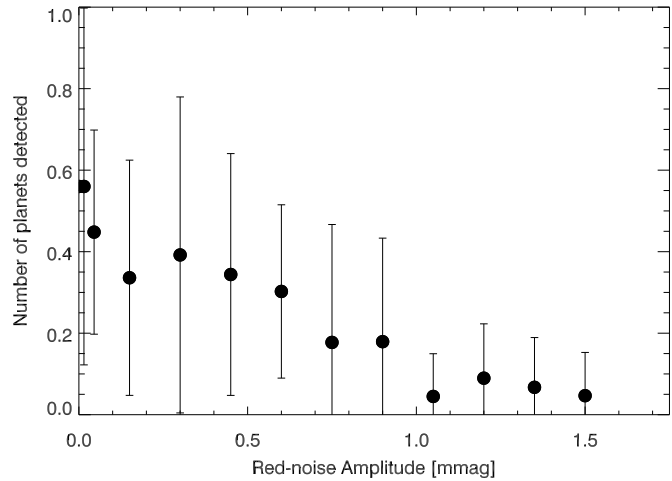}
    \caption[Red noise Dependency]{
      \label{RNDep}
      \label{lastfig}			
      Number of transit detection after two years of observation
      by a dedicated array of meter-class telescopes,
      as a function of red noise amplitude. Scaled to 
      an occurence fraction of 28\%.
    }
  \end{center}
\end{figure}

\end{document}